\documentclass[12pt,aps,showpacs]{revtex4}
\usepackage{graphicx}
\usepackage{amssymb}
\usepackage{amsmath}
\usepackage{bm}
\begin{document}
\title{$\alpha$-PARTICLE SPECTRUM IN THE REACTION p+$^{11}$B$\rightarrow \alpha + ^8Be^*\rightarrow 3\alpha$}
\author{V.~F.~Dmitriev}
\email{dmitriev@inp.nsk.su}
\affiliation{Budker Institute of Nuclear Physics,\\
 pr-t. Lavrentieva 11, Novosibirsk-90, 630090 \\
 and\\
 Novosibirsk State University,
 Pirogova 2, Novosibirsk-90, 630090}
\begin{abstract}
Using a simple phenomenological parametrization of the reaction amplitude we calculated $\alpha$-particle spectrum in the reaction p+$^{11}$B$\rightarrow \alpha + ^8Be^*\rightarrow 3\alpha$ at the resonance proton energy 675 KeV. The parametrization includes Breit-Wigner factor with an energy dependent width for intermediate $^8Be^*$ state and the Coulomb and the centrifugal factors in $\alpha$-particle emission vertexes. The shape of the spectrum consists of a well defined peak corresponding to emission of the primary $\alpha$ and a flat shoulder going down to very low energy. We found that below 1.5 MeV there are 17.5\% of $\alpha$'s and below 1 MeV there are 11\% of them.
\end{abstract}
\maketitle

\section{Introduction}
Successful realization of aneutronic fusion idea would greatly reduce problems associated with neutron radiation such as ionizing damage, neutron activation, and requirements for biological shielding, remote handling, and safety issues. One of the reactions discussed in this respect is the reaction p+$^{11}$B$\rightarrow \alpha + ^8Be^*\rightarrow 3\alpha$. There are at least two proposals \cite{rost,volos} how to use this reaction for energy production. The shape of $\alpha$-particles  spectrum, especially its low energy tail is of crucial importance for realisation of these projects.
Existing measurements \cite{beck,lin} produce the spectra for fixed angles of $\alpha$-particles that does not say much about integrated spectrum. In this paper we propose a simple theoretical model for the  reaction amplitude allowing to calculate with some assumptions the shape of $\alpha$-particle spectrum.

\section{Structure of the reaction amplitude}
The reaction p+$^{11}$B$\rightarrow 3\alpha$ has a well pronounced resonance at the proton energy 675 KeV.  The resonance corresponds to excited state of $^{12}C$ nucleus with the energy 16.57 MeV, and with quantum numbers $2^-$ and isospin $T=1$. The dominant decay mode is $\alpha$-particle plus excited state of $^8Be$ with the energy $E^*= 3.06$ MeV and quantum numbers $2^+$ and $T=0$. The cross section for decay into the ground state of $^8Be$ is about two orders of magnitude lower \cite{spit}. The decay proceeds with isospin violation, therefore the width of the resonance is small enough. With the $Q$-value $Q_1=6.14$ MeV the total width is $\Gamma=0.2$ MeV. It leads to high cross section at the resonance peak $\sim 1.2$ b. \cite{beck}.
The decaying state has negative parity, therefore the primary emitted $\alpha$ must be in the state with an odd angular momentum. In our case it can be $L=1$, and $L=3$.

 The width of the excited $^8Be$ state is 1.513 MeV. It corresponds to lifetime $4.35\times 10^{-22}$ s. During this time the primary $\alpha$-particle is still within strong interaction range. It means that the decay is in fact a 3-body type. Immediate consequence of it is that the energy of any $\alpha$-particle can vary from 0 to $2Q/3$, where $Q$ is total $Q$-value for the reaction, $Q =M_C+16.57-3M_\alpha= 9.295$ MeV. 

Keeping in mind these features of the reaction the amplitude can be presented in the following general form:
\begin{equation} \label{1}
T_1=V_2(\mathbf{p}_2+\mathbf{p}_1/2)\,G(-\mathbf{p}_1)\,V_1(\mathbf{p}_1),
\end{equation}
where $V_1(\mathbf{p}_1)$ is the emission vertex of the primary $\alpha$-particle with the momentum $\mathbf{p}_1$, $G$ is the propagator of the intermediate system $\alpha$-particle plus $^8Be^*$, and $V_2(\mathbf{p}_2+\mathbf{p}_1/2)$ is the decay vertex of $^8Be^*$ depending on relative momentum of two $\alpha$-particles. The diagram corresponding to this process is shown in Fig.1.
\begin{figure}[ht]
\includegraphics[width=5cm]{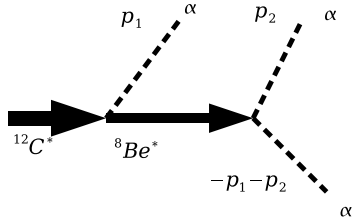}
\caption{Diagram of the reaction. Dashed lines correspond to emitted $\alpha$-particles.}
\end{figure}

For unpolarized initial state an angular distribution of $\alpha$-particle is isotropic. For this reason we omit angular dependence in the vertexes $ V_1(\mathbf{p})$ and $V_2(\mathbf{p})$ and will discuss only their energy dependence. There are two factors determining momentum dependence of the vertexes at low momentum. First, it is a Coulomb barrier that suppresses emission of low energy particles. Second, it is a centrifugal barrier. With these factors the decay vertex of $^8Be^*$ can be presented as
\begin{equation} \label{2}
V_2(p)=g_2\frac{p^2}{\sqrt{\kappa_2^4 +p^4}}\exp{(-2\pi\alpha/v)}\Gamma(1+\imath 4\alpha/v),
\end{equation}
where $v$ is a relative velocity of $\alpha$-particles in c.m. frame, and $\Gamma(x)$ is the Euler Gamma-function. The value of the cut-off parameter $\kappa_2$ is $\kappa_2 \sim 1/R$, where $R$ is the nuclear radius. The shape of the $\alpha$-particle spectrum appeared insensitive to exact value of $\kappa_2$. With this vertex we can define the energy dependent width by equation
\begin{equation} \label{3}
\Gamma_2(\epsilon)=g_2^2\frac{\alpha m_\alpha^2}{\exp(4\pi\alpha/v)-1}\frac{p^4}{\kappa_2^4+p^4}.
\end{equation}
 Here $\alpha$ is the fine structure constant, and $p=\sqrt{m_\alpha \epsilon}$, where $m_\alpha$ is the mass of $\alpha$-particle, and $\epsilon$ is the c.m. energy.
Remaining parameter $g_2$ is determined by the width of the excited state of $^8Be$
\begin{equation} \label{4}
\Gamma_2(\epsilon^*)=1.513,
\end{equation}
where $\epsilon^*=3.06$ MeV.
The vertex $V_1(p)$ has similar structure
\begin{equation} \label{5}
V_1(p)=g_1\frac{p}{\sqrt{\kappa_1^2 +p^2}}\exp{(-4\pi\alpha/v)}\Gamma(1+\imath 8\alpha/v).
\end{equation}
Here the Coulomb barrier is larger, $Z_1\times Z_2=8$. As for centrifugal barrier, strictly speaking there are two partial waves with $L=1$ and $L=3$ contributing to this vertex. However, since we are interested in low energy tail of the $\alpha$'s spectrum, we retained only $L=1$. In addition, the total width of the excited carbon state is small enough, $\Gamma_1=0.2$ MeV, therefore, the momentum dependence is smooth in this energy range and does not influence the shape of the spectrum. As in the previous vertex the exact value of the cut-off parameter $\kappa_1$ is not important. The shape of the spectrum is insensitive to it. The parameter $g_1$ is defined by normalization of the spectrum to one.

A propagator $G$ of the intermediate $\alpha + ^8Be^*$ state is of the form
\begin{equation} \label{6}
\hat{G}=\frac{1}{E-\hat{H}},
\end{equation}
where $E$ is the energy of the initial Carbon state $E=M_C+16.57$ MeV, and $\hat{H}$ is the Hamiltonian
of $\alpha + ^8Be^*$ system. Omitting $\alpha - ^8Be^*$ interaction and retaining only the width of $Be$ state we obtain
\begin{equation} \label{7}
G(-{\bm p}_1)=\frac{1}{M_C+16.57-3.06-M_{Be}-\epsilon_{ p_1}-E_{p_1}+\imath \Gamma_2(\epsilon)/2}.
\end{equation}
Here $\epsilon_p$ and $E_p$ are the laboratory kinetic energies of $\alpha$-particle and $^8Be$. Introducing $Q$-value for this transition we obtain,
\begin{equation} \label{8}
G(-{\bm p}_1)=\frac{1}{Q_1-1.5\epsilon_{{\bm p}_1}+\imath \Gamma_2(\epsilon_2)/2},
\end{equation}
where we used the relation $E_p=0.5\epsilon_p$. The center of mass energy $\epsilon$ in $\Gamma_2(\epsilon)$ is defined by total $Q$-value minus kinetic energy of $\alpha$ and $^8Be$, $\epsilon=Q-1.5\epsilon (p_1)$.

In the final state we have three identical Bose-particles. Therefore, the amplitude eq.(\ref{1}) should be properly symmetrized. The full amplitude is as follows
\begin{equation} \label{9}
T=\frac{V_2(\mathbf{p}_2+\mathbf{p}_1/2)\,V_1(\mathbf{p}_1)}{Q_1-1.5\epsilon_{ p_1}+\imath \Gamma_2(\epsilon_2)/2}+\frac{V_2(\mathbf{p}_1+\mathbf{p}_2/2)\,V_1(\mathbf{p}_2)}{Q_1-1.5\epsilon_{ p_2}+\imath \Gamma_2(\epsilon_1)/2}+\frac{V_2(\mathbf{p}_2/2-\mathbf{p}_1/2)\,V_1(-\mathbf{p}_1-\mathbf{p}_2)}{Q_1-1.5\epsilon_{{\bm p}_1+{\bm p}_2}+\imath \Gamma_2(\epsilon_{21})/2}.
\end{equation}
\section{Spectrum of the $\alpha$-particles}
With the amplitude eq.(\ref{9}) we can calculate the spectrum of $\alpha$-particles.
\begin{equation} \label{10}
dW=2\pi \delta (Q-\epsilon_{p_1}-\epsilon_{p_2} -\epsilon_{\mathbf{p}_1+\mathbf{p}_2})\overline{|T|^2}\frac{d^3p_1\,d^3p_2}{(2\pi)^6}.
\end{equation}
As usual, the line over the amplitude squared means averaging over initial spin projections and summation over angular momentum projections of $\alpha$-particles. After that the spectrum becomes isotropic. Strictly speaking, the interference terms in the amplitude squared may depend on the angle between $\mathbf{p}_1$ and $\mathbf{p}_2$. However, the scalar product $(\mathbf{p}_1\cdot\mathbf{p}_2)$ is fixed by delta-function in eq.(\ref{10}). The delta-function in eq.(\ref{10}) means that we neglected the small width of the initial excited Carbon state. Integration over angles removes delta-function, we obtain
\begin{equation} \label{11}
dW=\overline{|T|^2}\frac{m_\alpha p_1dp_1\,p_2dp_2}{4\pi^3}=\overline{|T|^2}\frac{m_\alpha^3 d\epsilon_1\,d\epsilon_2}{4\pi^3}.
\end{equation}
Integrating eq.(\ref{11}) over $\epsilon_2$ we obtain the final equation for $\alpha$-particle spectrum $dW/dE$
\begin{equation} \label{12}
\frac{dW}{dE}=A \int_{\epsilon_{min}}^{\epsilon_{max}} \overline{|T|^2} d\epsilon_2,
\end{equation}
where $\epsilon_{min}=(\sqrt{Q/2-3E/4}-\sqrt{E}/2)^2$, and $\epsilon_{max}=(\sqrt{Q/2-3E/4}+\sqrt{E}/2)^2$. The normalization factor $A$ defined by the condition
$$
\int_0^{2Q/3}\frac{dW}{dE}dE=1.
$$
The calculated spectrum is shown in Fig.2. The spectrum has a well pronounced peak corresponding to emission of primary $\alpha$-particle and a broad shoulder going down to very low energy. This low energy tail contains 11\% of $\alpha$-particles with the energy less than 1 MeV, and 17,5\% of $\alpha$-particles with the energy less than 1.5 MeV.
\begin{figure}[ht]
\includegraphics[width=12cm]{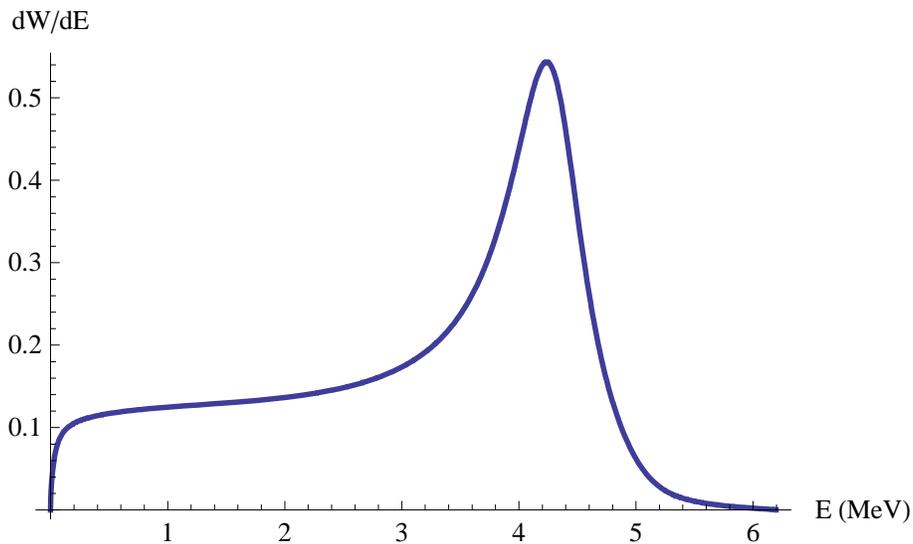}
\caption{Normalized to one the spectrum of $\alpha$-particles }
\end{figure}

It is worth to see how important are the Coulomb and the centrifugal factors included in the vertexes in eq.(\ref{2}). To do this we calculated the spectrum using constant $\Gamma_2=1.513$ MeV and constant $V_1$ and $V_2$. In Fig.3 we show the spectrum obtained in this way (dashed line) and the full spectrum from Fig.2.
\begin{figure}[ht]
\includegraphics[width=12cm]{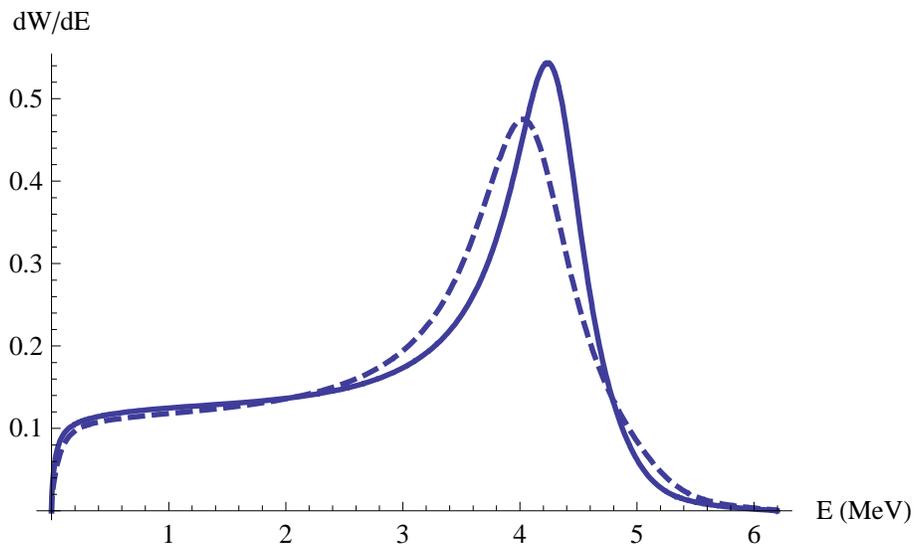}
\caption{Comparison of two spectra obtained for energy dependent (full line) and energy independent (dashed line) $\Gamma_2$ and the vertexes }
\end{figure}
The main difference is in the peak position which is shifted downward for $\sim 0.5$ MeV and an increase of number of $\alpha$-particles with the energies between 2.5 and 3.8 MeV. Note that in this region all three $\alpha$-particles have the energies close to each other. In this region the effect of interference of three terms in eq.(\ref{9}) is strong. The low energy tail is almost insensitive to it.

In summary, using a simple phenomenological model for the reaction amplitude the shape of the $\alpha$-particle spectrum has been calculated. Basing on the obtained shape we estimated the number of low energy $\alpha$'s. We found 17.5\% of  $\alpha$'s below 1.5 MeV and 11\% below 1 MeV.
\begin{acknowledgments}
Discussions with V.I. Volosov are greatly appreciated. This work was supported by RFBR grant 08-02-01155-a.
\end{acknowledgments}


\begin{thebibliography}{99}
\bibitem{rost}Norman Rostoker, Michl W. Binderbauer, Hendrik J. Monkhorst, Science, \textbf{278},1419 (1997).
\bibitem{volos} V.I. Volosov, Nucl. Fusion, \textbf{46}, 820 (2006).
\bibitem{beck} H.~Becker, C.~Rolfs, and H.~Trautwetter, Z. Phys., \textbf{A327}, 341 (1987).
\bibitem{lin} LIN Erh-kang {\it et} al., Chin.Phys.Lett., \textbf{15}, 796 (1998).
\bibitem{spit} C. Spitaleri {\it et} al., Phis.Rev. C \textbf{69}, 055806 (2004)/
\end{thebibliography}
\end{document}